\documentclass[twocolumn,nofootinbib,superscriptaddress]{revtex4-2}

\usepackage[dvipsnames]{xcolor}
\usepackage{hyperref}
\hypersetup{
  breaklinks=true,
  colorlinks=true,
  allcolors=BlueViolet,
}

\usepackage{graphicx} 
\graphicspath{{./figures/}} 

\usepackage{physics,braket,bm,amssymb} 
\usepackage{mathtools} 

\renewcommand{\t}{\text} 
\newcommand{\f}[2]{\dfrac{#1}{#2}} 
\newcommand{\p}[1]{\left(#1\right)} 
\renewcommand{\sp}[1]{\left[#1\right]} 
\renewcommand{\set}[1]{\left\{#1\right\}} 
\newcommand{\bk}{\Braket} 
\renewcommand{\v}{\bm} 
\newcommand{\B}{\mathcal{B}}
\newcommand{\E}{\mathcal{E}}
\newcommand{\F}{\mathcal{F}}
\newcommand{\I}{\mathcal{I}}
\renewcommand{\P}{\mathcal{P}}
\newcommand{\ZZ}{\mathbb{Z}}
\renewcommand{\i}{\text{i}}
\renewcommand{\o}{\text{o}}
\renewcommand{\c}{\text{c}}

\newcommand{\bbk}[1]{\langle\!\langle #1 \rangle\!\rangle}
\newcommand{\Bbk}[1]
{\left\langle\!\!\left\langle #1 \right\rangle\!\!\right\rangle}

\usepackage[inline]{enumitem} 
\setlist[enumerate]{label={(\roman*)}} 

\usepackage{orcidlink} 

\newcommand{\JILA}{JILA, National Institute of Standards and Technology and University of Colorado, 440 UCB, Boulder, Colorado 80309, USA}
\newcommand{\ALCF}{Argonne Leadership Computing Facility, Argonne National Laboratory, 9700 S.~Cass Avenue, Lemont, Illinois 60439}
\newcommand{\MCS}{Mathematics and Computer Science Division, Argonne National Laboratory, 9700 S.~Cass Avenue, Lemont, Illinois 60439}
\newcommand{\CSD}{Computational Science Division, Argonne National Laboratory, 9700 S.~Cass Avenue, Lemont, Illinois 60439}

\begin{document}

\title{Quantum Circuit Cutting with Maximum Likelihood Tomography}

\author{Michael A.~Perlin \orcidlink{0000-0002-9316-1596}}
\email{mika.perlin@gmail.com}
\affiliation{\JILA}
\affiliation{\ALCF}
\author{Zain H.~Saleem \orcidlink{0000-0002-8182-2764}}
\email{zsaleem@anl.gov}
\author{Martin Suchara \orcidlink{0000-0001-8808-1367}}
\email{msuchara@anl.gov}
\affiliation{\MCS}
\author{James C.~Osborn \orcidlink{0000-0001-7843-7622}}
\email{osborn@alcf.anl.gov}
\affiliation{\ALCF}
\affiliation{\CSD}

\begin{abstract}
We introduce maximum likelihood fragment tomography (MLFT) as an improved circuit cutting technique for running clustered quantum circuits on quantum devices with a limited number of qubits.
In addition to minimizing the classical computing overhead of circuit cutting methods, MLFT finds the most likely probability distribution for the output of a quantum circuit, given the measurement data obtained from the circuit's fragments.
We demonstrate the benefits of MLFT for accurately estimating the output of a fragmented quantum circuit with numerical experiments on random unitary circuits.
Finally, we show that circuit cutting can estimate the output of a clustered circuit with higher fidelity than full circuit execution, thereby motivating the use of circuit cutting as a standard tool for running clustered circuits on quantum hardware.
\end{abstract}

\maketitle

\section{Introduction}

The advent of noisy intermediate-scale quantum (NISQ) technologies \cite{preskill2018quantum} makes quantum processors with increasing numbers of qubits available to the quantum computing community for experimentation.
The rapid progress in the development and manufacturing of these devices is remarkable, with state-of-the-art superconducting quantum processors reaching $\sim50$ qubits with percent-level gate and readout errors \cite{wendin2017quantum, kjaergaard2020superconducting, arute2019quantum}.
Advances on the hardware front have been matched by the theoretical development of suitable hardware benchmarks \cite{boixo2018characterizing}, which have in turn enabled proof-of-principle demonstrations of a computational advantage over classical computing systems \cite{arute2019quantum}.

Despite tremendous progress, existing devices still lack the number and quality of qubits required for practical NISQ-era applications such as digital quantum simulation \cite{lloyd1996universal, georgescu2014quantuma}, quantum optimization \cite{farhi2014quantum, hadfield2019quantum, moll2018quantum} and quantum machine learning \cite{dunjko2016quantumenhanced, biamonte2017quantum}.
Without error correction, these applications are severely limited by the accumulation of errors that will only compound as devices scale up to more qubits and deeper circuits.
Bridging the gap between the requirements of NISQ-era quantum algorithms and the capabilities of NISQ devices will require error mitigation techniques \cite{endo2018practical, kandala2019error} and problem decompositions that trade quantum and classical computing resources \cite{bravyi2016trading, peng2020simulating}.

One decomposition, inspired by the fragmentation methods used for quantum molecular cluster simulations \cite{li2007generalized, li2008fragmentationbased, gordon2012fragmentation}, applies fragmentation to the execution of quantum circuits \cite{peng2020simulating}.
This decomposition consists of first ``cutting'' a quantum circuit into smaller subcircuits, or ``fragments'', that can be executed on processors with fewer qubits, and then reconstructing the probability distribution over measurement outcomes for the original quantum circuit from probability distributions associated with its fragments.
The severed quantum connections between circuit fragments are simulated by classical post-processing of fragment data, which leads to a classical computing overhead that grows exponentially with the number of cuts that are made to a circuit.
This approach is therefore suitable for simulating circuits that are decomposable into clusters of gates with a small number of inter-cluster interactions.
Such circuits can make appearances in the context of Hamiltonian simulation \cite{peng2020simulating}, as well as near-term applications based on a variational ansatz that allows for some freedom in choosing circuit structure, such as the quantum approximate optimization algorithm (also the quantum alternating operator ansatz, QAOA) \cite{farhi2014quantum, hadfield2019quantum, saleem2020approaches, saleem2021scaling} and variational quantum eigensolvers (VQE) \cite{peruzzo2014variational, peng2020simulating}.

Due to the presence of fundamental shot noise (equivalently, finite sampling error), an unavoidable feature of the original fragment recombination method in Ref.~[\citenum{peng2020simulating}] is that the distribution over measurement outcomes obtained by characterizing and recombining circuit fragments does not generally satisfy central axioms of probability theory, namely that a probability distribution must be non-negative and normalized.
A naive fix to this problem would be to simply remove all negative probabilities and normalize the reconstructed distribution in question.
In the spirit of maximum likelihood state tomography (MLST) \cite{smolin2012efficient}, however, one would like to determine the ``most likely'' probability distribution that is consistent with available fragment data.

In this work, we find this ``most likely'' probability distribution by generalizing MLST and introducing maximum likelihood fragment tomography (MLFT), the use of which guarantees that reconstructed probability distributions are non-negative and normalized.
We discuss how MLFT minimizes the classical computing resources necessary to characterize circuit fragments, and provide a tensor-network-based method for fragment recombination.
We test our methods in numerical experiments with random unitary circuits, and demonstrate that MLFT estimates the probability distribution at the output of a fragmented quantum circuit with higher fidelity than the naive method of removing negative probabilities and normalizing.
These benefits come at no cost to the computational complexity of circuit cutting, as they are achieved by post-processing fragment data in a manner that has a smaller computational cost than that of recombining fragment data to reconstruct a circuit output.
As an added bonus, for a fixed number of queries to quantum hardware (known as ``shots'' or ``trials'' in e.g.~Qiskit \cite{qiskit} or pyQuil \cite{smith2017practical}) we show that circuit cutting methods can outperform direct execution and sampling of a clustered circuit in order to estimate its associated probability distribution.
We provide theoretical arguments to support this finding, which motivates the use of circuit cutting as a standard tool for evaluating clustered circuits on quantum hardware, even when all hardware requirements for full circuit execution are satisfied.

\section{Background}

Here we provide a basic overview and discussion of the circuit cutting procedure first introduced in Ref.~[\citenum{peng2020simulating}], and establish terminology that we will use throughout the rest of this work.
We note that our overview of circuit cutting will use the language of quantum states and channels, rather than the language of tensor networks that was used in Ref.~[\citenum{peng2020simulating}].
These two formalisms are mathematically equivalent, but the former will allow for a more seamless integration with the material in Section \ref{sec:MLFT}.

\subsection{The general cut-and-stitch prescription}

\begin{figure*}
\centering
\includegraphics{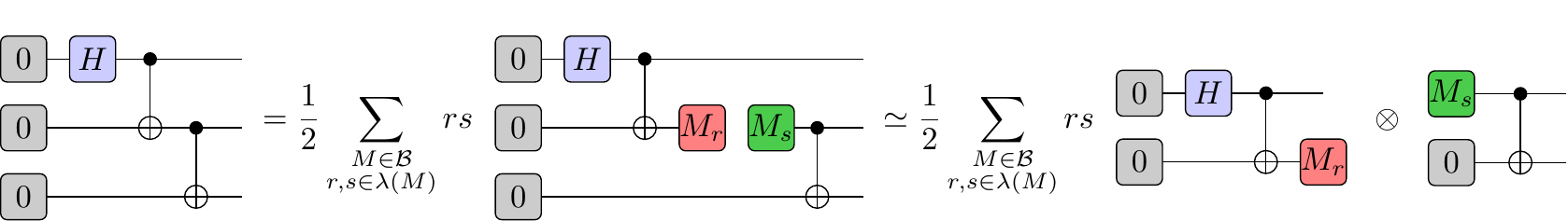}
\caption{
{\bf Circuit cutting example.}
A 3-qubit GHZ circuit can be cut into two 2-qubit fragments by inserting an identity operator.
Here $\B\equiv\set{X,Y,Z,I}$ is the set of Pauli operators $X,Y,Z$ and the identity $I$, which together form an orthogonal basis for the space of single-qubit operators; $\lambda\p{M}$ denotes the spectrum of $M$; and $M_s\equiv\op{M_s}$ is the projector onto an eigenstate $\ket{M_s}$ of $M$ with eigenvalue $s$.
Green (red) boxes labeled by the state $M_s$ ($M_r$) correspond to preparations (projections) of a qubit in the corresponding state.
After cutting a circuit, the resulting fragments can be simulated independently, and an appropriate post-processing of simulation results recovers the output of the original (pre-cut) circuit.
}
\label{fig:cut_example}
\end{figure*}

Given an arbitrary quantum state $\ket\psi$ of $N$ qubits, a straightforward resolution of the identity operator $I=\sum_{b\in\set{0,1}}\op{b}$ on qubit $n$ implies that
\begin{align}
  \ket\psi = I_n \ket\psi
  \simeq \sum_{b\in\set{0,1}}
  \ket{b} \otimes \prescript{}{n}{\bk{b|\psi}},
  \label{eq:pure_identity}
\end{align}
where $I_n$ denotes the action of $I$ on qubit $n$; the relation $\simeq$ denotes equality up to a permutation of tensor factors (i.e.~qubit order); and $\prescript{}{n}{\bk{b|\psi}}$ is a sub-normalized state of $N-1$ qubits acquired by projecting $\ket\psi$ onto state $\ket{b}$ of qubit $n$.
If the structure of a quantum circuit that prepares $\ket\psi$ allows, a similar resolution of the identity operator $I$ can be used to ``cut'' the circuit by inserting $I$ at a location that splits the circuit into two disjoint subcircuits.
For example, if $\ket\psi=V_{23}U_{12}\ket{000}$, where $U_{12}$ and $V_{23}$ are the two-qubit gates $U$ and $V$ acting on qubits $1,2$ and $2,3$, then by inserting the identity operator $I_2$ (on qubit 2) between $U_{12}$ and $V_{23}$ we find that
\begin{align}
  \ket\psi \simeq \sum_{b\in\set{0,1}} \ket{\psi_1\p{b}} \otimes \ket{\psi_2\p{b}},
  \label{eq:pure_fragments}
\end{align}
where the factors
\begin{align}
  \ket{\psi_1\p{b}} \equiv \prescript{}{2}{\bk{b|U|00}},
  &&
  \ket{\psi_2\p{b}} \equiv V \ket{b0}
\end{align}
are (generally sub-normalized) ``conditional'' states prepared by projecting onto $\ket{b}$ or preparing $\ket{b}$, as appropriate.
The identity in Eq.~\eqref{eq:pure_fragments} is visualized in Figure \ref{fig:cut_example}, albeit with the use of density operators that we discuss below.

The above splitting method relies on the capability to project qubit $n$ onto state $\ket{b}$ while preserving phase information.
Such capability is possible when running classical simulations of a circuit, but is not possible on quantum computing hardware.
This limitation can be overcome by representing quantum states $\ket\psi$ with density operators $\rho=\op\psi$, whose diagonal entries in a given measurement basis define a classical probability distribution over measurement outcomes in that basis.
For ease of language, we will at times blur the distinction between a state $\rho$ and the probability distribution defined by its diagonal entries in a fixed computational basis.
In the remainder of this work, we will discuss circuit splitting and reconstruction in way that is compatible with circuit execution on quantum computing hardware.
Nonetheless, our methods can be applied just as well to classical state simulation, with minor simplifying modifications to account for the added capability of performing deterministic, phase-preserving qubit projections.

The identity analogous to Eq.~\eqref{eq:pure_identity} for density operators $\rho$ reads
\begin{align}
  \rho \simeq \f12 \sum_{M\in\B} M \otimes \tr_n\p{M_n\rho},
  \label{eq:cut_identity}
\end{align}
where $\B$ is a basis of self-adjoint $2\times 2$ matrices with normalization $\tr\sp{M^{(i)} M^{(j)}} = 2 \delta_{ij}$ for $M^{(i)},M^{(j)}\in\B$; $\tr_n$ denotes a partial trace with respect to qubit $n$; and $M_n$ with $n\in\ZZ_n$ denotes an operator that acts with $M$ on qubit $n$ and trivially (i.e.~with the identity $I$) on all other qubits.
To be concrete, we will use the set of Pauli operators together with the singe-qubit identity operator, $\B\equiv\set{X,Y,Z,I}$, as our basis.
The identity in Eq.~\eqref{eq:cut_identity} implies that the state prepared by the action of a three-qubit circuit $V_{23}U_{12}$ on the trivial state $\op{0}^{\otimes 3}$ can be decomposed as
\begin{align}
  \rho \simeq \f12 \sum_{M\in\B} \rho_1\p{M} \otimes \rho_2\p{M},
  \label{eq:cut_fragments}
\end{align}
where now the factors
\begin{align}
  \begin{split}
    \rho_1\p{M} &\equiv \tr_2\p{M_2 U \op{0}^{\otimes 2} U^\dag}, \\
    \rho_2\p{M} &\equiv V \p{M \otimes \op{0}} V^\dag,
  \end{split}
\end{align}
have no straightforward interpretation as ``conditional'' states, as with $\ket{\psi_1\p{b}}$ and $\ket{\psi_2\p{b}}$ in Eq.~\eqref{eq:pure_fragments}.
In order to decompose $\rho$ into conditional states, we can expand each $M\in\B$ in its eigenbasis:
\begin{align}
  \rho \simeq \f12 \sum_{\substack{M\in\B\\r,s\in\lambda\p{M}}}
  r s \, \rho_1\p{M_r} \otimes \rho_2\p{M_s},
  \label{eq:cut_fragments_states}
\end{align}
where $\lambda\p{M}$ denotes the spectrum of $M$, i.e.~$\lambda\p{X}=\lambda\p{Y}=\lambda\p{Z}=\p{+1,-1}$ and $\lambda\p{I}=\p{1,1}$; and $M_s\equiv\op{M_s}$ with $s\in\lambda\p{M}$ is a projector onto an eigenstate of $\ket{M_s}$ of $M$ with eigenvalue $s$.
Note that the choice of eigenstates for the identity operator $I$ is arbitrary as long as these two states are orthogonal, so we can reuse the eigenstates from one of the other operators.

The decomposition in Eq.~\eqref{eq:cut_fragments_states} allows interpreting each $\rho_f\p{M_s}$ as a conditional state, obtained either by post-selecting onto the measurement of a qubit in state $\ket{M_s}$, or by preparing a qubit in state $\ket{M_s}$, as appropriate (see Figure \ref{fig:cut_example}).
This decomposition thus corresponds to the following procedure for circuit cutting and reconstruction: after cutting a circuit into (say) two fragments, characterize the classical probability distributions $\rho_f\p{M_s}$ over measurement outcomes by running the corresponding sub-circuit and either post-selecting on measurement outcomes $M_s$ or preparing states $M_s$, as appropriate.
Note that post-selected probability distributions are generally sub-normalized, and the normalization $\tr\rho_1\p{M_s}$ is equal to the probability of getting outcome $M_s$ when measuring in the diagonal basis of $M$.
After characterizing the conditional distributions $\rho_f\p{M_s}$ for each of $f\in\set{1,2}$, $M\in\set{X,Y,Z}$, and $s\in\set{+1,-1}$, combine these distributions according to Eq.~\eqref{eq:cut_fragments_states}.
This scenario is illustrated in Figure \ref{fig:cut_example}, which cuts a 3-qubit GHZ circuit preparing the state $\ket\psi\propto\ket{000}+\ket{111}$ into two 2-qubit fragments.

\subsection{Refinements}

In practice, recombining circuit fragments as prescribed by Eq.~\eqref{eq:cut_fragments_states} is inefficient in two ways.
First, the tensor products in Eq.~\eqref{eq:cut_fragments_states} are a computational bottleneck for fragment recombination.
It is therefore faster to post-process conditional distributions by first
\begin{enumerate*}
\item for each fragment $f$, combining the six independent distributions $\rho_f\p{M_s}$ into four distributions: $\rho_f\p{M}=\rho_f\p{M_{+1}}-\rho_f\p{M_{-1}}$ for each $M\in\set{X,Y,Z}$ and $\rho_f\p{I}=\rho_f\p{M_{+1}}+\rho_f\p{M_{-1}}$ for any $M\in\set{X,Y,Z}$, and then
\item combining the fragment distributions $\rho_f\p{M}$ according to Eq.~\eqref{eq:cut_fragments}.
\end{enumerate*}
In a circuit with $K$ cuts, this post-processing reduces the number of tensor products that must be computed during recombination from $16^K$ to $4^K$, which is an exponential reduction (in $K$) of the number of floating-point operations required to recombine fragment data\footnote{
The recombination procedure in Ref.~[\citenum{peng2020simulating}] involves $8^K$ tensor products, rather than $16^K$, because it consolidates ``measurement'' conditions, but not ``preparation'' conditions, which is equivalent to collapsing the sum over $r$ in Eq.~\eqref{eq:cut_fragments_states} but leaving the sum over $s$.
}.

Second, the recombination formula in Eq.~\eqref{eq:cut_fragments_states} nominally requires, for each fragment $f$ incident on $K_f$ cuts, characterizing $K_f^6$ probability distributions.
This characterization is overcomplete, because the $K_f^6$ distributions are not all linearly independent.
In the case of a fragment with a single incident cut, for example, we can use the fact that $X_++X_-=Z_++Z_-=I$ to decompose
\begin{align}
  \rho_f\p{X_-} = \rho_f\p{Z_+} + \rho_f\p{Z_-} - \rho_f\p{X_+}.
  \label{eq:linear_dependence}
\end{align}
In fact, a fragment with $K_f$ incident cuts can be completely characterized by $K_f^4$ distributions, which can be deduced from the fact that the space of operators on the Hilbert space of a qubit has real dimension four.
The symmetric, informationally complete, positive operator-valued measure (SIC-POVM) $\set{\Pi^{\t{SIC}}_j:j\in\ZZ_4}$ (consisting of projectors $\Pi^{\t{SIC}}_j$ onto the states represented by the four corners of a regular tetrahedron inscribed in a Bloch sphere), for example, form a mutually unbiased basis for the space of single-qubit operators.
Given any single-qubit operator $M$, we can therefore expand $\rho_f\p{M}=\sum_{j\in\ZZ_4}c_{Mj}^{(f)}\rho_f\p{\Pi^{\t{SIC}}_j}$ with real coefficients $c_{Mj}^{(f)}$.

Finally, characterizing fragments is a noisy process, due to both
\begin{enumerate*}
\item hardware errors that are unavoidable without error correction, as in all NISQ devices, and
\item statistical sampling (shot) noise.
\end{enumerate*}
As a result, the ``experimentally inferred'' distributions $\tilde\rho_f\p{M_s}$ approximating the ``true'' distributions $\rho_f\p{M_s}$ will generally contain errors, and will fail to satisfy self-consistency conditions such as Eq.~\eqref{eq:linear_dependence}.
When combining these distributions according to Eqs.~\eqref{eq:cut_fragments} and \eqref{eq:cut_fragments_states} there is similarly no guarantee that the reconstructed probability distribution will satisfy conditions required of a probability distribution, such as non-negativity and normalization.

To address these shortcomings, in the following section we recast the task of characterizing conditional distributions into the task of performing fragment tomography, treating the fragments $\rho_f$, rather than distributions $\rho_f\p{M_s}$, as first-class objects.
In addition to being automatically efficient in terms of the classical memory footprint of characterizing each fragment, performing fragment tomography allows us to adapt the method of maximum likelihood tomography \cite{smolin2012efficient} to construct a model for each fragment that is, by construction, guaranteed to satisfy all appropriate self-consistency conditions.
Fragment recombination is then similarly guaranteed to yield a probability distribution that is both non-negative and normalized.
Finally, we show how the fragment models constructed via fragment tomography naturally admit a tensor-network-based method for recombination.

\section{Maximum likelihood fragment tomography}
\label{sec:MLFT}

Once a circuit has been cut into fragments $\rho_f$, rather than characterizing conditional distributions $\rho_f\p{M_s}$ we can perform a more systematic maximum likelihood fragment tomography (MLFT) procedure to characterize these fragments.
The purpose of MLFT is to perform a ``maximum likelihood'' characterization, similar to the characterization of quantum states in Ref.~[\citenum{smolin2012efficient}], which guarantees that any probability distribution associated with these fragments will be (i) the ``most likely'' distribution consistent with available fragment data, while (ii) satisfying all necessary constraints for a valid (i.e.~non-negative and normalized) probability distribution.
MLFT is a type of quantum process tomography, which generalizes maximum likelihood state tomography (MLST) \cite{smolin2012efficient} to the case of channels (processes) with mixed (quantum/classical) inputs and outputs.

\begin{figure*}
\centering
\includegraphics{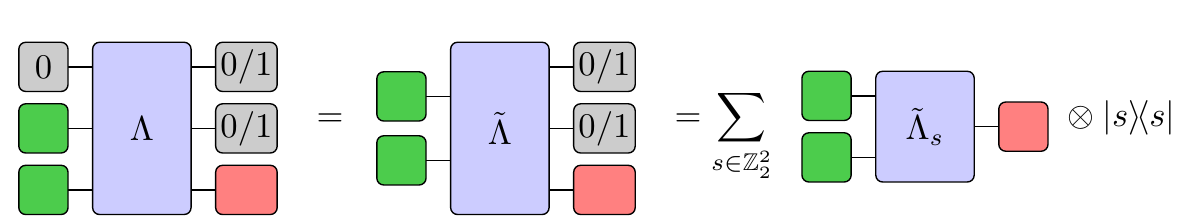}
\caption{
{\bf Block-diagonalizing circuit fragments.}
Each circuit fragment can be identified with a density operator $\Lambda$ on the joint Hilbert space of its input (left) and output (right) qubits.
Classical inputs and outputs of a fragment (gray) correspond to qubits that are either prepared in the trivial state $\ket{0}$ (labeled ``$0$'') or measured in a fixed computational basis (labeled ``$0/1$'').
Quantum inputs (left, green) and outputs (right, red) correspond to qubits associated with cuts in a circuit.
Due to the presence of trivial inputs, we only need to characterize a reduced state $\tilde\Lambda$ on the Hilbert space of the quantum inputs and all outputs.
Classical outputs give this reduced state a block-diagonal structure: $\tilde\Lambda=\sum_s\tilde\Lambda_s\otimes\op{s}$, where the block $\tilde\Lambda_s$ is associated with the measurement of bitstring $s$ on the classical outputs of the fragment.
}
\label{fig:MLFT_sketch}
\end{figure*}

Any given fragment, nominally a unitary circuit on $Q$ qubits, will generally have $Q_\i$ ``quantum input'' and $Q_\o$ ``quantum output'' qubits at the locations of cuts.
We refer to these inputs and outputs as ``quantum'' because characterizing the fragment for circuit reconstruction will require performing full quantum tomography on the corresponding degrees of freedom.
In contrast, the remaining $C_\i\equiv Q-Q_\i$ ``classical input'' qubits are always initialized in the trivial state $\ket{\bm 0}_\i\equiv\ket{0}^{\otimes C_\i}$, and the remaining $C_\o\equiv Q-Q_\o$ ``classical output'' qubits are always measured in a fixed computational basis.
For definiteness, we can first think of a fragment as a quantum channel $\E_\Lambda$ on the state of $Q$ qubits.
The channel-state duality \cite{jamiolkowski1972linear, choi1975completely, jiang2013channelstate} implies that this channel is uniquely determined by a 4-partite state (density operator) of the form
\begin{align}
 \Lambda \equiv
  \sum_{\substack{k,\ell,m,n\\p,q,r,s}}
  \Lambda_{k\ell;mn;pq;rs} \op{k}{\ell} \otimes \op{m}{n}
  \otimes \op{p}{q} \otimes \op{r}{s},
\end{align}
where the bitstrings $k,\ell$ ($m,n$; $p,q$; $r,s$) index states in the Hilbert space of the quantum input (classical input; quantum output; classical output) qubits of the fragment, and are implicitly summed over $\ZZ_2^{Q_\i}$ ($\ZZ_2^{C_\i}$; $\ZZ_2^{Q_\o}$; $\ZZ_2^{C_\o}$).
Specifically, the channel $\E_\Lambda$ maps a bipartite input state
\begin{align}
  \rho \otimes \op{\bm 0_\i}
  \equiv \sum_{k,\ell} \rho_{k\ell} \op{k}{\ell} \otimes \op{\bm 0_\i}
\end{align}
at its input to the bipartite state
\begin{align}
  \E_\Lambda\p{\rho\otimes\op{\bm 0_\i}}
  \equiv
  \sum_{k,\ell,p,q,r,s}
  \Lambda_{k\ell;0,0;pq;rs} \rho_{k\ell} \op{p}{q} \otimes \op{r}{s}
\end{align}
at its output.
To account for the fact that classical outputs are only ever measured in a fixed computational basis, we can remove all parts of $\E_\Lambda\p{\rho}$ that are off-diagonal with respect to the measurement basis of the corresponding qubits.
In total, we therefore need only characterize the channel $\E_{\tilde\Lambda}$ defined by
\begin{align}
  \E_{\tilde\Lambda}\p{\rho}
  \equiv \sum_{k,\ell,p,q,s}
  \tilde\Lambda_{k\ell;pq;s} \rho_{k\ell} \op{p}{q} \otimes \op{s},
\end{align}
where
\begin{align}
  \tilde\Lambda_{k\ell;pq;s} \equiv \Lambda_{k\ell;0,0;pq;ss}\,.
\end{align}
The task of performing MLFT thus reduces to performing tomography on the tri-partite block-diagonal state
\begin{align}
  \begin{split}
    \tilde\Lambda
    &\equiv \sum_{k,\ell,p,q,s}\tilde\Lambda_{k\ell;pq;s}\op{k}{\ell}\otimes\op{p}{q}\otimes\op{s} \\
    &= \sum_s \tilde\Lambda_s \otimes \op{s},
  \end{split}
  \label{eq:blocks}
\end{align}
where Eq.~\eqref{eq:blocks} implicitly defines the blocks $\tilde\Lambda_s$.
In words, the reduced state $\tilde\Lambda$ is acquired from the full state $\Lambda$ by conditioning on (i.e.~fixing) a trivial state $\op{\bm 0_\i}$ on its classical inputs, and the block $\tilde\Lambda_s$ is acquired from $\tilde\Lambda$ by conditioning on measurement of the bitstring $s$ on its classical outputs.
The relationship between $\Lambda$, $\tilde\Lambda$, and $\tilde\Lambda_s$ is sketched out in Figure \ref{fig:MLFT_sketch}.

In a nutshell, MLFT is performed by providing a variety of quantum inputs to $\E_{\tilde\Lambda}$, and measuring its quantum outputs in a variety of bases.
The blocks $\tilde\Lambda_s$ are inferred by least-squares fitting to a linear operator that maps quantum inputs to quantum outputs, using all available data from experiments in which bitstring $s$ was observed on the classical outputs of a fragment.
This procedure yields an experimental ansatz state $\Lambda_{\t{A}}$ that approximates $\tilde\Lambda$, but that generally does not have the properties required of a density operator, such as a non-negative spectrum.
The last step in MLFT is therefore to convert the ansatz state $\Lambda_{\t{A}}$ into a ``maximum likelihood'' state $\Lambda_{\t{ML}}$ by using an algorithm borrowed from MLST in Ref.~[\citenum{smolin2012efficient}].
We describe MLFT in more detail below.

MLFT (and MLST) begins by collecting measurement data to characterize the quantum state under consideration.
In the case of the block-diagonal state $\tilde\Lambda$, one needs to characterize the expectation values
\begin{align}
  \bk{\sigma_\i\otimes\sigma_\o\otimes z_\c}_{\tilde\Lambda}
  \equiv \tr\sp{\tilde\Lambda\p{\sigma_\i\otimes\sigma_\o\otimes z_\c}}
\end{align}
for some complete basis of operators $\set{\sigma_\i\otimes\sigma_\o\otimes z_\c}$ on the target Hilbert space of $\tilde\Lambda$, where $\sigma_\i$, $\sigma_\o$, and $z_\c$ are respectively operators on the quantum input, quantum output, and classical output of the fragment in question, with $z_\c$ strictly diagonal in the computational basis.
MLST \cite{smolin2012efficient} collects data by performing informationally complete measurements of $\tilde\Lambda$, for example by choosing operators $\sigma_{\i,\o}$ from the set of all Pauli strings $\set{I,X,Y,Z}^{\otimes Q_{\i,\o}}$, and choosing $z_\c$ from the set of diagonal Pauli strings $\set{I,Z}^{\otimes C_\o}$.
In the case of fragment tomography, however, we do not have direct access to the state $\tilde\Lambda$, and instead have access to the channel $\E_{\tilde\Lambda}$.
It is therefore not possible to directly measure the degrees of freedom in $\tilde\Lambda$ that are associated with inputs to the channel.
Instead, MLFT characterizes the quantum input degrees of freedom in $\tilde\Lambda$ by preparing an informationally complete set of states, making use of the fact that
\begin{align}
  \begin{split}
    \tr\sp{\tilde\Lambda\p{\sigma_\i\otimes\sigma_\o\otimes z_\c}}
    &= \tr\sp{\E_{\tilde\Lambda}\p{\sigma_\i^{\t{T}}}\p{\sigma_\o\otimes z_\c}} \\
    &= \bk{\sigma_\o\otimes z_\c}_{\E_{\tilde\Lambda}\p{\sigma_\i^{\t{T}}}},
  \end{split}
\end{align}
where $\sigma_\i^{\t{T}}$ denotes the transpose of $\sigma_\i$.
Whereas the operators $\sigma_\o$ and $z_\c$ may still be chosen from the set of Pauli strings, the input state $\sigma_\i^{\t{T}}$ is restricted to satisfy $\tr\sigma_\i^{\t{T}}=1$.
This restriction excludes the possibility of choosing $\sigma_\i^{\t{T}}$ from an orthogonal basis for the space of the space of $Q_\i$-qubit operators (such as the set of Pauli strings), but any complete basis will suffice.
For example, one can choose input states from the basis of pure states $\set{\ket{0},\ket{1},\ket{0}+\ket{1},\ket{0}+i\ket{1}}^{\otimes Q_\i}$.
For an unbiased basis, one can take tensor products of symmetric informationally complete (SIC) states of a single qubit, or even consider bases of multi-qubit SIC states.
The practical advantages of using these bases, however, generally depend on the fidelity with which one can prepare SIC states.
Similar considerations apply for the choice of measurement basis for quantum outputs \cite{adamson2010improving}.
Overall, in order to characterize a fragment with $Q_\i$ quantum inputs and $Q_\o$ quantum outputs one must prepare each of $4^{Q_\i}$ input states, and measure outputs in each of $3^{Q_\o}$ possible bases (for each quantum output qubit, the diagonal bases of $X,Y,Z$), so fragment tomography requires $O\p{4^{Q_\i}3^{Q_\o}}$ experiments.

\begin{figure*}
\centering
\includegraphics{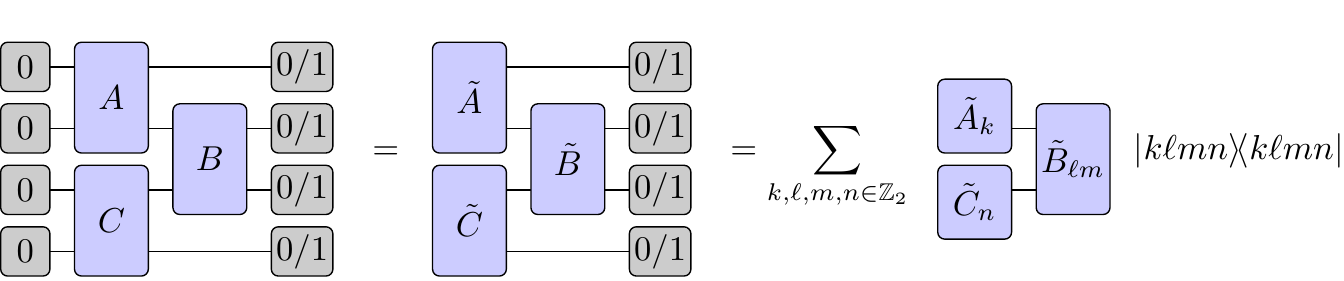}
\caption{
{\bf Fragment recombination as a tensor network contraction problem.}
The full probability distribution over measurement outcomes for a circuit reconstructed from fragments $A,B,C$ can be represented by a tensor contraction of the reduced states $\tilde A,\tilde B,\tilde C$, obtained by performing MLFT on the fragments.
The probability to measure a given bitstring $k\ell mn$ (i.e.~a concatenation of $k,\ell,m,n\in\ZZ_2$) on the output of the fragment is given by the contraction of the diagonal blocks $\tilde A_k,\tilde B_{\ell m},\tilde C_n$.
The lack of classical inputs to fragment $B$ implies that $\tilde B=B$.
}
\label{fig:recombination}
\end{figure*}

After collecting an informationally complete set of data on the state $\tilde\Lambda$, a straightforward least-squares fitting procedure yields an empirical ansatz $\Lambda_{\t{A}}$ for $\tilde\Lambda$, which is the MLFT analogue of the ``experimentally noisy'' matrix $\mu$ described in the original MLST work \cite{smolin2012efficient}.
The block diagonal structure of $\tilde\Lambda=\sum_s\tilde\Lambda_s\otimes\op{s}$
implies that the least-squares fitting procedure can be performed independently for each block $\tilde\Lambda_s$ of size $2^{Q_\i+Q_\o}\times2^{Q_\i+Q_\o}$.
Specifically, $\tilde\Lambda_s$ is obtained by fitting to
\begin{align}
  \tr\sp{\tilde\Lambda_s\p{\sigma_\i\otimes\sigma_\o}}
  = p_s \bk{\sigma_\i\otimes\sigma_\o}_{z_\c=s},
\end{align}
where $p_s$ is the probability of observing bitstring $s$ on the classical output of a fragment, and $\bk{\sigma_\i\otimes\sigma_\o}_{z_\c=s}$ is the expectation value of $\sigma_\o$ (on the quantum outputs) when preparing the state $\sigma_\i^{\t{T}}$ (on the quantum inputs) and observing bitstring $s$ (on the classical outputs) of the fragment.
Because the ansatz state $\Lambda_{\t{A}}\approx\tilde\Lambda$ is constructed from a fit to noisy measurement data, $\Lambda_{\t{A}}$ will generally have negative eigenvalues, which is not allowed for density operators.
The final step in both MLST and MLFT is therefore to find the closest state to $\Lambda_{\t{A}}$ that has no negative eigenvalues.
To this end, MLFT borrows the ``fast algorithm for subproblem 1'' in Ref.~[\citenum{smolin2012efficient}], which
\begin{enumerate}
  \item diagonalizes $\Lambda_{\t{A}}$,
  \item eliminates the most negative eigenvalue (setting it to zero),
  \item adds an equal amount to all other eigenvalues to enforce $\tr\Lambda_{\t{A}}=1$, and
  \item repeats steps (ii,iii) until there are no more negative eigenvalues.
\end{enumerate}
As proven in Ref.~[\citenum{smolin2012efficient}], this algorithm finds the closest positive semidefinite state $\Lambda_{\t{ML}}$ to $\Lambda_{\t{A}}$ with respect to the metric induced by the 2-norm $\norm{A}_2\equiv\sqrt{\tr\p{A^\dag A}}$.
In this sense, $\Lambda_{\t{ML}}$ is the ``most likely'' state consistent with $\Lambda_{\t{A}}$.
The only additional consideration for this algorithm when performing MLFT has to do with making use of block diagonal structure to diagonalize $\Lambda_{\t{A}}$: each block of size $2^{Q_\i+Q_\o}\times2^{Q_\i+Q_\o}$ can be diagonalized independently.
The overall serial runtime of the algorithm to find $\Lambda_{\t{ML}}$ from $\Lambda_{\t{A}}$ is therefore $O\p{2^{3(Q_\i+Q_\o)}N_\c}$, where $N_\c\le 2^{C_\o}$ is the number of blocks in $\Lambda_{\t{A}}$, or equivalently the number of distinct bitstrings observed on the classical output of the fragment throughout tomography.
As we will see, the maximum-likelihood corrections to $\Lambda_{\t{A}}$ are responsible for the benefits of MLFT in estimating a circuit's output.
Moreover, the cost of computing these corrections is smaller than the unavoidable cost of fragment recombination, so the benefits of MLFT are free as far as the computational complexity of circuit cutting is concerned.

The treatment of fragments and their dual states $\Lambda$ as first-class objects in MLFT enables a straightforward tensor-network-based circuit reconstruction method.
Rather than explicitly computing and summing over each term of the fragment recombination formula in Eq.~\eqref{eq:cut_identity}, the basic idea is to think of the entire sum as a contraction of two tensors.
We sketch out this idea in Figure \ref{fig:recombination}, making use of the relationship between fragment states $\Lambda$, their reductions $\tilde\Lambda$, and diagonal blocks $\tilde\Lambda_s$.
In total, the full probability distribution over measurement outcomes for a reconstructed circuit can be acquired by a tensor network contraction of reduced states $\tilde\Lambda$, and the individual probabilities of measuring any given bitstring at the output of a circuit can be acquired by a similar contraction of diagonal blocks $\tilde\Lambda_s$.

If a circuit has $K$ cuts and $F$ fragments, and $N_\c^{(f)}$ distinct bitstrings were observed on the classical output of fragment $f\in\set{1,2,\cdots,F}$ throughout fragment tomography, then reconstructing the circuit's output requires contracting $\prod_f N_\c^{(f)}$ tensor networks, each of which nominally involves summing over $4^K$ terms.
Whereas the $4^K$ cost to contract a single tensor network $g$ can be reduced to $2^{O(\t{cc}(g))}$, where $\t{cc}(g)$ is the contraction complexity of $g$ \cite{peng2020simulating}, the overall multiplicative cost in $N_\c^{(f)}$ is unavoidable.
In comparison, performing maximum-likelihood corrections to fragment models comes at a cost that is additive in $N_\c^{(f)}$.
For this reason, fragment recombination is generally the computational bottleneck of circuit cutting, and maximum-likelihood corrections add no significant overhead.

\section{Numerical experiments}

\begin{figure}
\centering
\includegraphics{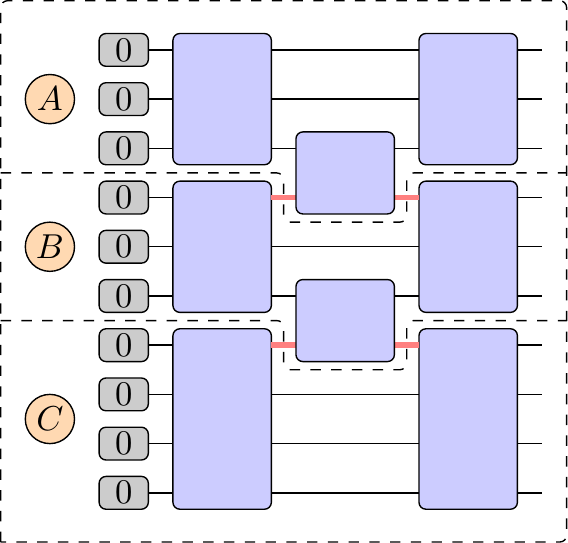}
\caption{
{\bf Random unitary circuit (RUC) of ten qubits split into three clusters.}
Qubits are first split among clusters as evenly as possible, and each cluster is prepared in a random state by the application of a Haar-random unitary gate \cite{zyczkowski1994random, emerson2005convergence}.
Adjacent clusters are then entangled with random two-qubit gates, before again applying a layer of random unitaries on all clusters.
A clustered RUC is cut into fragments (labeled $A,B,C$) by cutting the bottom legs (shown in red) of every inter-cluster entangling gate.
}
\label{fig:clustered_circuit}
\end{figure}

In order to test the benefits of MLFT in an application-agnostic setting, we run classical simulations of random unitary circuits (RUCs).
Because the cost of circuit cutting scales exponentially with the number of cuts made to a circuit, we construct RUCs with a structure that makes them amenable to circuit cutting (see Figure \ref{fig:clustered_circuit}).
We then vary the number of qubits and clusters in our RUCs, as well as the total number of samples (known as ``shots'' in Qiskit \cite{qiskit} or ``trials'' in pyQuil \cite{smith2017practical}) in a simulation, where the result of each sample is a single bitstring representing one measurement outcome.
In this way, we compare three methods to estimate the probability distribution over measurement outcomes at the end of a clustered RUC.

First, as a standard benchmark, we consider sampling an entire circuit $S$ times without any circuit cutting, which we refer to as the method of ``full'' circuit execution.
Second, we consider cutting a circuit into fragments, with each fragment corresponding to a cluster as shown in Figure \ref{fig:clustered_circuit}, and reconstructing these fragments as prescribed by the original circuit cutting work \cite{peng2020simulating}, namely without maximum likelihood corrections.
We refer to this second method as the ``direct'' method of circuit cutting and reconstruction.
A fragment with $Q_\i$ quantum inputs and $Q_\o$ quantum outputs has $4^{Q_\i}\times 3^{Q_\o}$ variants that must be simulated for circuit reconstruction, where each variant corresponds to a choice of state preparations and measurement bases on the quantum inputs and outputs of the fragment.
We therefore divide the budget of $S$ samples evenly among all fragment variants.
Finally, we consider the full MLFT and recombination procedure, which we refer to as the ``MLFT'' method.
The direct and MLFT methods only differ in the classical post-processing of fragment simulation results.
Specifically, the differences between the final outputs of the direct and MLFT methods are entirely due to the application (or non-application) of maximum-likelihood corrections to fragment models.

To compare the efficacy of the full, direct, and MLFT methods, we compute the fidelity of reconstructed probability distributions over measurement outcomes, $p_{\t{estimate}}$, with the actual probability distribution $p_{\t{actual}}$ that is determined by exact classical simulations of a circuit:
\begin{align}
  \F = \sp{\sum_s \sqrt{p_{\t{actual}}\p{s} \, p_{\t{estimate}}\p{s} }}^2,
\end{align}
where $p_{\t{actual}}\p{s}, p_{\t{estimate}}\p{s}$ are, respectively, the probabilities of measuring the $N$-qubit state (bitstring) $s\in\ZZ_2^N$ according to the distributions $p_{\t{actual}}, p_{\t{estimate}}$.
The fidelity $\F$ is an analogue of the quantum state overlap $\abs{\bk{\phi|\psi}}^2$ for classical probability distributions.
The only caveat in our calculation of fidelities is that they are only well defined when dealing with valid (non-negative and normalized) probability distributions, whereas the direct circuit cutting method generally yields an unnormalized distribution that may have negative entries.
We therefore convert the distribution yielded by the direct method into a valid probability distribution by eliminating all negative entries (setting them to zero), and normalizing the distribution.

\begin{figure*}
\centering
\includegraphics{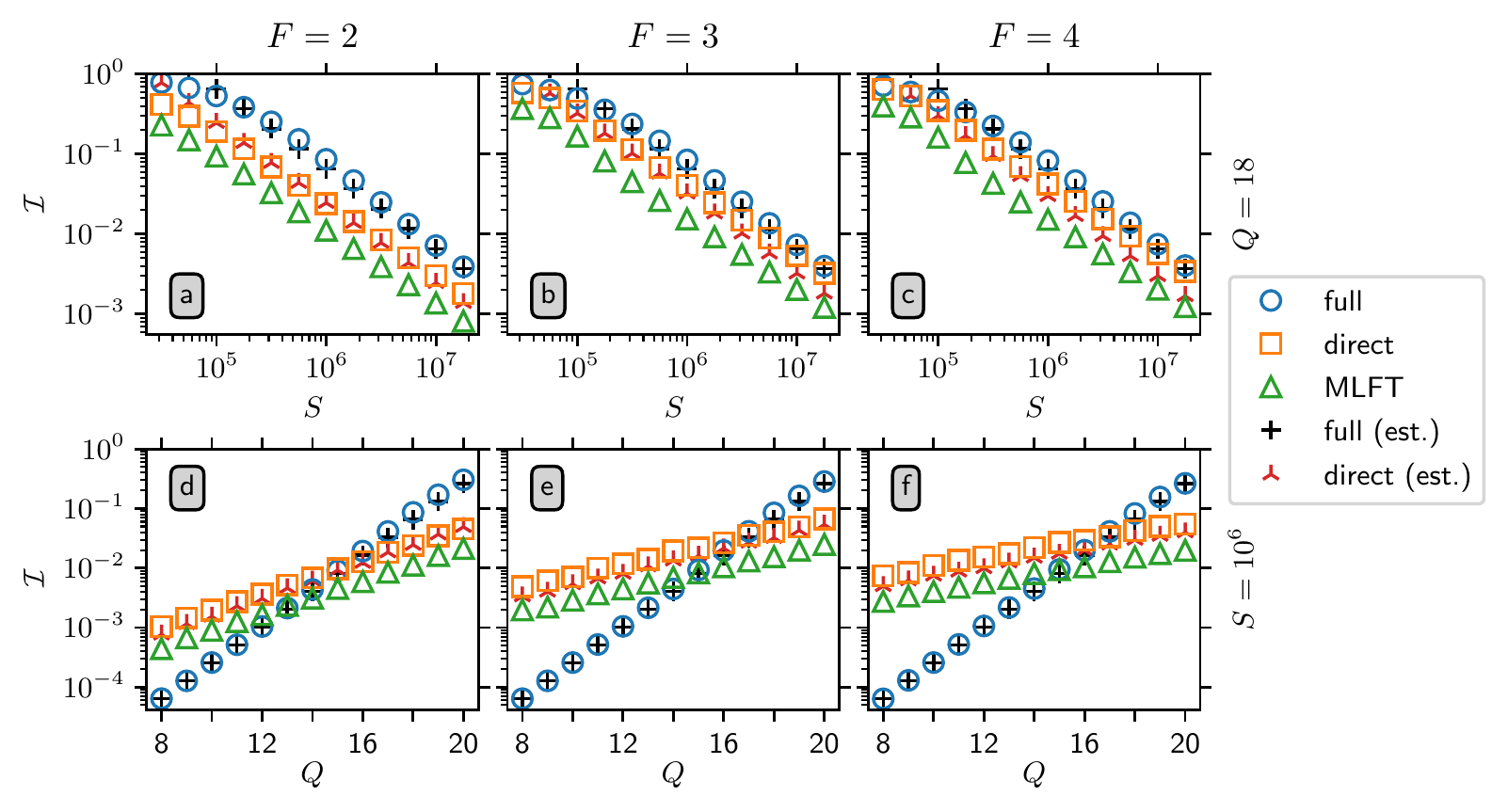}
\caption{
{\bf Infidelity in reconstructed circuit outputs.}
The infidelity $\I=1-\F$ as a function of sample number $S$ ({\bf a, b, c}) or qubit number $Q$ ({\bf d, e, f}) for clustered random unitary circuits (RUCs) with $F=2$ ({\bf a, d}), $3$ ({\bf b, e}) or $4$ ({\bf c, f}) fragments.
Open markers correspond to simulations of the full circuit (``full''), or simulations via circuit cutting before (``direct'') and after (``MLFT'') maximum likelihood corrections to fragment models.
The last two markers in the legend correspond to analytical estimates of infidelity: $2^Q/S$ for the full method, and $\sum_{f=1}^F 2^{C^f_\o}/n$ for the direct method, where $C^f_\o$ is the number of classical outputs on fragment $f$ and $n=S/V$ is the number of samples devoted to each of $V$ total fragment variants.
Whereas the estimates for the full method are quantitatively accurate, the estimates for the direct method are provided only to highlight approximate scaling relationships (see Appendices \ref{sec:preliminaries}--\ref{sec:infidelity_cut}).
Results for each data point are averaged over 100 instances of a clustered RUC.
}
\label{fig:infidelities}
\end{figure*}

Figure \ref{fig:infidelities} shows the infidelities $\I=1-\F$ of the probability distributions yielded by each simulation method.
To ensure that results are not sensitive to the specific choice of random gates, these infidelities are averaged over 100 instances of each clustered RUC, although in practice we find that these infidelities vary by only $\sim1$--$10$\% of their mean value (see Appendix \ref{sec:infidelities_std}).
Figure \ref{fig:infidelities} also shows analytical estimates of infidelity for the full and direct simulation methods, derived in Appendices \ref{sec:preliminaries}--\ref{sec:infidelity_cut}.

An immediate takeaway from Figure \ref{fig:infidelities} is that the MLFT method introduced in this work always outperforms the direct method: MLFT infidelities are always lower than direct infidelities.
This result is consistent with theoretical arguments that MLFT finds the ``most likely'' fragment model consistent with noisy measurement data.
Although we only consider shot noise in this work, it would be interesting to see how the benefits of MLFT change with the introduction of additional noise such as measurement and gate errors.
We defer a study of the effect of such errors to future work.

Figure \ref{fig:infidelities} also shows that the infidelity $\I$ for all simulation methods scales more or less identically with the sample number $S$, namely $\I\sim1/S$ for large $S$.
Though some of the numerical data in Figure \ref{fig:infidelities} may better be fit by $\I\propto1/S^{(1+\eta)}$ for some $\eta\ne0$, the deviation from $\eta=0$ are minor, and may be an artifact of small circuit sizes.
It is worth noting that the original circuit cutting work \cite{peng2020simulating} proved that a reconstructed circuit output (probability distribution) can be estimated to an accuracy of $\epsilon$ with $S=O(1/\epsilon^2)$ samples, which by dimensional analysis suggests that $\I\sim\epsilon^2\sim1/S$ in all cases.

Though scaling with sample number does not strongly distinguish these methods, it is clear that the direct and MLFT methods scale much more favorably with circuit size: the full method has an infidelity $\I\sim 2^Q$ for $Q$ qubits, whereas cutting a circuit into $F$ fragments results in $\I\sim\sum_{f=1}^F 2^{C^f_\o}$, where $C^f_\o\approx Q/F$ is the number of classical outputs on fragment $f$ and $\sum_f C^f_\o=Q$.
The more favorable scaling for circuit cutting methods is surprising at first glance, as these methods require strictly fewer quantum computing resources: their sample budget is spent on executing smaller circuits (namely, fragment variants).
The better performance of the circuit cutting methods can be understood by the fact that they use their sample budget in a targeted manner that exploits circuit structure, rather than blindly sampling the entire circuit.
However, when circuits are sufficiently small for the fixed number of samples to explore the sample space of the entire circuit, full circuit sampling performs better than circuit cutting because it does not waste resources on characterizing numerous variants of nearly identical fragments.

Deferring a detailed derivation of expected infidelities to Appendices \ref{sec:preliminaries}--\ref{sec:infidelity_cut}, we can make the above intuition more quantitative by considering the difficulty of estimating a probability distribution defined by a $Q$-qubit RUC by (i) sampling the full circuit directly, versus (ii) sampling all fragment variants for circuit reconstruction.
The first task requires, in principle, exploring a sample space of size $2^Q$ with $S$ samples, so one might reasonably expect (as is indeed the case) that $\I\sim 2^Q/S$.
If a circuit is cut into $F$ fragments, meanwhile, then each fragment will have $\sim Q/F$ qubits, and if the number of qubits is independent of the number of fragments, then the overall sample space volume is reduced from $2^Q$ to $2^{O(Q/F)}$.
Indeed, this argument agrees with the estimate of infidelity for the direct method of circuit cutting in Figure \ref{fig:infidelities}, where we show that $\I\sim F\times 2^{Q/F}/n$ with $n=S/V$ the number of samples devoted to each of $V$ total fragment variants.

\section{Conclusions and outlook}

Circuit cutting is a promising technique for reducing the qubit requirements of running clustered quantum circuits.
We have introduced improved circuit cutting methods by minimizing associated classical computing costs (with an exponential improvement over previous methods), and by using MLFT to reconstruct the ``most likely'' probability distribution defined by a quantum circuit, given the measurement data obtained from its fragments.
To test our ideas in an application-agnostic setting, we ran classical simulations of random unitary circuits, which demonstrate the advantages of MLFT compared to the original circuit cutting method.
Moreover, we also show that circuit cutting has advantages as a standard technique for running clustered circuits on quantum hardware, even when full circuit execution is possible.

Our work opens several avenues for the improvement and application of circuit cutting techniques.
For example, MLFT guarantees that fragment models satisfy appropriate self-consistency conditions, but MLFT makes no use of the fact that each fragment corresponds to a unitary quantum channel.
Furthermore, our present work neglects the effects of hardware errors that are important to consider in the context of NISQ devices.
Because MLFT has the capability to mitigate shot noise, we expect the advantages of MLFT over full circuit execution to be enhanced when additionally considering the effects of hardware errors.
We likewise expect unitarity constraints to provide additional benefits for mitigating sources of noise.
Our work thus complements ongoing efforts that study the benefits of circuit cutting in the presence of hardware errors, which have generally found that circuit cutting helps mitigate the effects of noise  \cite{ayral2020quantum}.
Having framed fragment characterization as a tomography task, it would also be interesting to adapt and apply different quantum process tomography techniques \cite{torlai2020quantum} to the task of circuit cutting, and compare their performance and cost to that of MLFT.

As a final point, we note that circuit cutting in its current form estimates a probability distribution associated with a given circuit.
Ideally, one would like to sample this probability distribution (defined over an exponentially large space of possible measurement outcomes) without having to reconstruct it in full.
To this end, our work makes important progress in understanding the mechanics of circuit cutting, by providing a convenient and efficient framework for thinking about individual circuit fragments.
We hope that this framework will help in achieving the ultimate the goal of sampling a quantum circuit by sampling its fragments.

\section*{Acknowledgments}

We acknowledge helpful discussions with Yuri Alexeev, Bradley Pearlman, Teague J. Tomesh, Wei Tang, Thomas Ayral, and Francois-Marie Le R\'{e}gent.
This material is based upon work supported by Laboratory Directed Research and Development (LDRD) funding from Argonne National Laboratory, provided by the Director, Office of Science, of the U.S. Department of Energy under contract DE-AC02-06CH11357; the Argonne Leadership Computing Facility, which is DOE Office of Science User Facility supported under Contract DE-AC02-06CH11357; and the National Science Foundation under Award No. 2037984.

\section*{Author Contributions}

M.S.~motivated the study of circuit cutting, and brought attention to the problem of unphysical (negative and unnormalized) reconstructed probability distributions.
J.C.O.~had the idea to build maximum likelihood fragment models.
Z.H.S.~thought to adapt existing maximum likelihood state tomography techniques to this task.
M.A.P.~worked out the theory, wrote the codes, and drafted the manuscript, with aid and consulting from Z.H.S.
M.S.~and~J.C.O.~supervised the project.
All authors discussed the numerical experiments, interpreted results, and provided critical feedback and contributions to the final manuscript.

\section*{Code and Data Availability}

The codes, circuits, and simulation data used for numerical experiments in this work are available in the online repository \verb|Quantum-Software-Tools/QSPLIT-MLFT| \cite{qsplit-mlft}.

\bibliography{main.bib}

\appendix
\onecolumngrid
\setcounter{figure}{0}
\renewcommand{\thefigure}{\thesection\arabic{figure}}

\section{Preliminaries}
\label{sec:preliminaries}

In these appendices, we derive
\begin{enumerate*}
\item the expected infidelity of a circuit output estimated using the ``full'' method in the main text (i.e.~full circuit execution and sampling), and
\item an asymptotic bound on the expected infidelity of a circuit output estimated using the ``direct'' method in the main text (i.e.~with circuit cutting, but without maximum-likelihood corrections to fragment models).
\end{enumerate*}
In practice, when comparing with numerical experiments we find that our asymptotic bound for the ``direct'' method is overly pessimistic in its scaling with the total number of cuts $K$ in a fragmented circuit.
Nonetheless, this upper bound provides a scaling with fragment size that agrees with numerical results.

\subsection{Multinomial distribution sampling error}

Let $p$ be a classical probability distribution over a discrete (and finite) set of measurement outcomes $\set{b}$, and let $p_b$ be the probability of outcome $b$.
We denote an empirical estimate of $p_b$ by $\tilde p_b$, and denote the error in this estimate by $\epsilon_b=\tilde p_b-p_b$.
If we take $n$ samples of $p$ and set each $\tilde p_b$ to the fraction of times that we observed outcome $b$, then the statistical means and covariances of the errors $\epsilon_b$ are
\begin{align}
  \bbk{\epsilon_b} = 0,
  &&
  \bbk{\epsilon_b\epsilon_c} = \f{p_b \p{\delta_{bc} - p_c}}{n},
  \label{eq:errors}
\end{align}
where $\bbk{X}$ denotes the expected value of $X$ after averaging over attempts to estimate $p$ from $n$ samples; and $\delta_{bc}=1$ if $b=c$ and zero otherwise.

\subsection{Sampling infidelity}

Let $p_b$ be the probability of observing bitstring $b$ at the end of a circuit, and $\tilde p_b=p_b+\epsilon_b$ an empirical estimate of $p_b$.
The infidelity of the estimated probability distribution is
\begin{align}
  \I = 1 - \sp{\sum_b \sqrt{p_b \tilde p_b}}^2
  = 1 - \sp{\sum_b p_b \sqrt{1 + \f{\epsilon_b}{p_b}}}^2,
\end{align}
where we tentatively assume that all $p_b\ne0$.
Expanding the square root as $\sqrt{1+x}\approx1+x/2-x^2/8+O(x^3)$, up to $O\p{\epsilon^3}$ corrections we find that
\begin{align}
  \I &\approx 1 - \sp{\sum_b p_b \p{1 + \f12\f{\epsilon_b}{p_b}
      - \f18\f{\epsilon_b^2}{p_b^2}}}^2 \\
  &= 1 - \sum_{b,c} p_b p_c \p{1 + \f12\f{\epsilon_b}{p_b}
    - \f18\f{\epsilon_b^2}{p_b^2}}
  \p{1 + \f12\f{\epsilon_c}{p_c}
    - \f18\f{\epsilon_c^2}{p_c^2}} \\
  &\approx 1 - \sum_{b,c} p_b p_c
  \sp{1 + \f12\p{\f{\epsilon_b}{p_b} + \f{\epsilon_c}{p_c}}
    + \f14\f{\epsilon_b\epsilon_c}{p_b p_c}
    - \f18 \p{\f{\epsilon_b^2}{p_b^2} + \f{\epsilon_c^2}{p_c^2}}} \\
  &= -\sum_b \epsilon_b
  - \f14 \sum_{b,c} \epsilon_b \epsilon_c
  + \f14 \sum_b \f{\epsilon_b^2}{p_b}.
  \label{eq:infidelity}
\end{align}

\section{Sampling infidelity with full circuit execution}
\label{sec:infidelity_full}

Let $p_b$ be the probability of observing bitstring $b$ at the end of the circuit, and $\tilde p_b=p_b+\epsilon_b$ an empirical estimate of $p_b$.
If we sample the probability distribution $n$ times and set each $\tilde p_b$ to the fraction of times that we observed bitstring $s$, then the estimates $\tilde p_b$ of $p_b$ are normalized with $\sum_b\tilde p_b=1$, so
\begin{align}
  \sum_b \epsilon_b = \sum_b \p{\tilde p_b - p_b}
  = \sum_b \tilde p_b - \sum_b p_b = 0.
\end{align}
Up to $O(\epsilon^3)$ corrections, the expected infidelity $\bbk{\I}$ is then
\begin{align}
  \bbk{\I}
  \approx \f14 \sum_b \f{\bbk{\epsilon_b^2}}{p_b}
  = \f14 \sum_b \f{1-p_b}{n}
  = \f{2^Q-1}{4n} \approx \f{2^Q}{4n},
  \label{eq:full_infidelity}
\end{align}
where $2^Q = \sum_b 1$ is the total number of bitstrings that can be measured at the output of $Q$ qubits.

A few comments concerning the result in Eq.~\eqref{eq:full_infidelity} are in order.
First, the restriction that $\I\in\sp{0,1}$ implies that Eq.~\eqref{eq:full_infidelity} can only hold for $4n>2^Q-1$, and that if $4n$ is comparable to $2^Q$ then the $O(\epsilon^3)$ contributions to $\I$ must become relevant.
Second, if any $p_b=0$, then the factor $2^Q$ should be replaced by the sample space volume $\abs{\set{b:p_b\ne0}}$.
This second observation in particular suggests that Eq.~\eqref{eq:full_infidelity} can only hold for sufficiently `generic' probability distributions, as large separations of scale in the probabilities $p_b$ should reduce $2^Q$ to some smaller `effective' sample space volume, likely determined by the output entropy $S(p)=-\sum_b p_b\log p_b$.
Finally, we point out that Eq.~\eqref{eq:full_infidelity} also describes the infidelity with which $n$ samples estimate the conditional probability distributions associated with a single fragment of a cut-up circuit.
Unfortunately, the presence of quantum correlations between circuit fragments implies that the infidelity of a reconstructed circuit output is not additive in the infidelities of the fragments.
Nonetheless, we show in the following section that the infidelity of a reconstructed circuit still scales inversely with the number of fragment samples, and exponentially in fragment size, i.e.
\begin{align}
  \bbk{\I} \sim O\p{\f1n \sum_f 2^{C^f_\o}},
\end{align}
where now $n$ is the number of samples used to estimate each variant of each fragment, $f$ indexes a single fragment, and $C^f_\o$ is the number of classical output bits on fragment $f$.

\section{Sampling infidelity with circuit cutting}
\label{sec:infidelity_cut}

We now attempt to compute the expected infidelity of a circuit output estimated from fragment data.
We first work out the relatively simple case of one cut and two fragments, which we will subsequently generalize to the case of arbitrary cuts and fragments.

\subsection{One cut, two fragments}

In order to evaluate the expected infidelity $\bbk{\I}$, we first need to construct an estimator $\tilde p$ of $p$.
We can take the diagonal elements of Eqs.~(5) and (7) in the main text (in the computational basis) to expand the probability distribution
\begin{align}
  p = \f12 \sum_{M\in\B} p^1\p{M} \otimes p^2\p{M}
  &&
  p^f\p{M} = \sum_{r\in\lambda\p{M}} r \, p^f\p{M_r},
\end{align}
where $p^f\p{M_r}$ is the probability distribution at the output of fragment $f$ conditional on measuring or initializing $M_r$, as appropriate.
We will assume without loss of generality that the first fragment is measured and that the second fragment is initialized at the cut, in which case $p^1\p{M_r}$ is normalized to the probability of getting outcome $M_r$ when measuring the appropriate qubit in the diagonal basis of $M$.
To deal with this normalization properly, we expand
\begin{align}
  p^f\p{M_r} = a^f\p{M_r} q^f\p{M_r},
\end{align}
where $a^f\p{M_r}=\sum_b p_b^f\p{M_r}$ is the normalization of $p^f\p{M_r}$ (which is equal to 1 for initialization conditions), and $q^f\p{M_r}= p^f\p{M_r}/a^f\p{M_r}$ is a normalized probability distribution.

We construct estimators $\tilde p^f\p{M_r}$ of $p^f\p{M_r}$ as follows.
For measurement conditions, the estimator $\tilde a^f\p{M_r}\approx a^f\p{M_r}$ is set to the fraction of times that we observe outcome $M_r$ when measuring in the appropriate basis; for initialization conditions, $\tilde a^f\p{M_r}=a^f\p{M_r}=1$.
Each entry $\tilde q^f_b\p{M_r}$ of the probability distribution $\tilde q^f\p{M_r}\approx q^f\p{M_r}$ is set to the fraction of times that we observe bitstring $b$ on the classical output of fragment $f$ when conditioned on $M_r$.
For any argument $M_r$ of $a^f$, $q^f$, etc., we can then define the errors $\beta^f=\tilde a^f-a^f$ and $\gamma^f=\tilde q^f-q^f$ and expand
\begin{align}
  \tilde p^f = \tilde a^f \tilde q^f = p^f + \epsilon^f,
  &&
  \epsilon^f = \beta^f q^f + a^f \gamma^f + \beta^f \gamma^f,
\end{align}
and define, for all symbols $X\in\set{p^f,a^f,q^f,\tilde p^f,\tilde a^f,\tilde q^f,\epsilon^f,\beta^f,\gamma^f}$,
\begin{align}
  X\p{M} = \sum_{s\in\lambda\p{M}} s \, X\p{M_s},
  \label{eq:M_cond}
\end{align}
which allows us to construct the estimator
\begin{align}
  \tilde p
  = \f12 \sum_{M\in\B}
  \tilde p^1\p{M} \otimes \tilde p^2\p{M}
  = p + \epsilon,
\end{align}
where
\begin{align}
  \epsilon = \f12 \sum_{M\in\B}
    \sp{\epsilon^1\p{M} \otimes p^2\p{M}
    + p^1\p{M} \otimes \epsilon^2\p{M}
    + \epsilon^1\p{M} \otimes \epsilon^2\p{M}}.
  \label{eq:eps_2}
\end{align}
Strictly speaking, the definition of estimators and errors $X\p{I}$ are ambiguous as presented, as there is not a unique decomposition of $I$ for the sum in Eq.~\eqref{eq:M_cond}.
In practice, our implementations of circuit cutting algorithms set
\begin{align}
  X\p{I} = \f13 \sum_{\substack{M\in\set{X,Y,Z}\\r\in\lambda\p{M}}}
  X\p{M_r}.
  \label{eq:I_cond}
\end{align}
Whereas the errors $\epsilon^f\p{M}$ and $\epsilon^f\p{M'}$ for $M,M'\in\set{X,Y,Z}$ are uncorrelated unless $M=M'$, the decomposition in Eq.~\eqref{eq:I_cond} implies that $\epsilon^f\p{I}$ is correlated with $\epsilon^f\p{M}$ for all $M\in\B$.
For simplicity, however, we will assume that the estimators $\tilde p^f\p{M}$ and $\tilde p^f\p{M'}$ with $M\ne M'$ are built from independent experimental data, such that their corresponding errors $\epsilon^f\p{M}$ and $\epsilon^f\p{M'}$ are uncorrelated.
Crucially, this assumption does not affect the general structure of our calculations, and therefore leaves our main conclusions (namely, how reconstruction infidelity $\bbk{\I}$ scales with different fragment parameters) in tact.

In order to compute the expected infidelity of $\tilde p$, we now need to determine the statistical means $\bbk{\epsilon_b}$, covariances $\bbk{\epsilon_b\epsilon_c}$, and variances $\bbk{\epsilon_b^2}$.
The means $\bbk{\epsilon_b}=0$ because all contributions to $\epsilon_b$ are either (i) proportional to a single error in the estimate of a multinomially distributed random variable, which is mean-zero, or (ii) a product of multiple independent (uncorrelated) errors, which is also mean-zero.

\subsubsection{Covariances}

To compute the covariance $\bbk{\epsilon_b\epsilon_c}$, we note that are only working to second order in error variables, whereas the contributions to $\bbk{\epsilon_b\epsilon_c}$ from the $\sim\epsilon^1\otimes\epsilon^2$ terms of Eq.~\eqref{eq:eps_2} are either fourth order or zero; we are therefore free to neglect these terms.
Additionally throwing out terms that vanish because they are the product of uncorrelated random variables, we find that
\begin{align}
  \bbk{\epsilon_b\epsilon_c}
  \approx \f14 \sum_{f,M}
  \bbk{\epsilon_{b_f}^f\p{M} \epsilon_{c_f}^f\p{M}} \,
  p_{b_{\bar f}}^{\bar f}\p{M} p_{c_{\bar f}}^{\bar f}\p{M},
  \label{eq:cov_full}
\end{align}
where $f\in\set{1,2}$ and $\bar f\ne f$, i.e.~such that $\set{f,\bar f}=\set{1,2}$; and $b_f,c_f$ are the substrings of $b,c$ associated with fragment $f$.
We now expand
\begin{align}
  \begin{split}
    \bbk{\epsilon_b^f\p{M} \epsilon_c^f\p{M}}
    &\approx \sum_{r,s\in\lambda\p{M}} r
    s \, \bbk{\beta^f\p{M_r} \beta^f\p{M_s}} \,
    q_b^f\p{M_r} q_c^f\p{M_s} \\
    &\quad + \sum_{r,s\in\lambda\p{M}}
    r s \, a^f\p{M_r} a^f\p{M_s}
    \bbk{\gamma_b^f\p{M_r} \gamma_c^f\p{M_s}},
  \end{split}
  \label{eq:cov_frag_start}
\end{align}
where we again throw out terms that are zero or fourth order in error variables.
The normalization errors $\beta^f\p{M_r},\beta^f\p{M_s}$ are zero for initialization conditions, and are always correlated for measurement conditions because they are errors in mutually exclusive measurement outcomes.
The probability distribution errors $\gamma_b^f\p{M_r},\gamma_c^f\p{M_s}$, meanwhile, are independent unless $M_r=M_s$.
The covariances between these errors are determined by multinomial distribution sampling errors, so
\begin{align}
  \bbk{\beta^f\p{M_r} \beta^f\p{M_s}}
  &= \f1n \, a^f\p{M_r} \sp{\delta_{rs} - a^f\p{M_s}}
  \p{1 - \delta_{Q^f_\o,0}},
  \label{eq:norm_error} \\
  \bbk{\gamma_b^f\p{M_r} \gamma_c^f\p{M_s}}
  &= \f{\delta_{rs}}{a^f\p{M_r} n} \,
  q^f_b\p{M_r} \sp{\delta_{bc} - q^f_c\p{M_r}},
  \label{eq:dist_error}
\end{align}
where $n$ is the number of times that we sample fragment each variant of each fragment (i.e.~each choice of initialization conditions and measurement bases on a fragment), $a^f\p{M_r} n$ is the expected number of times that we sample fragment $f$ with condition $M_r$, $Q^f_\o$ is the number of quantum outputs (i.e.~or measurement conditions) on fragment $f$.

If $\delta_{Q^f_\o,0}=0$, then the $\sim a^f\p{M_r} \delta_{rs}$ contributions from Eq.~\eqref{eq:norm_error} cancel out with the $\sim q^f_b\p{M_r} q^f_c\p{M_r}$ contributions from Eq.~\eqref{eq:dist_error} when substituting these results into Eq.~\eqref{eq:cov_frag_start}.
Meanwhile, if $\delta_{Q^f_\o,0}=1$ then $a^f\p{M_r}=1$, so
\begin{align}
  n\, \bbk{\epsilon_b^f\p{M} \epsilon_c^f\p{M}}
  \approx \delta_{bc} \, p^f_b\p{I}
  - \p{1-\delta_{Q^f_\o,0}} p^f_b\p{M} p^f_c\p{M}
  - \delta_{Q^f_\o,0} \sum_r p^f_b\p{M_r} p^f_c\p{M_r}.
  \label{eq:cov_frag}
\end{align}
Altogether,
\begin{align}
  \sum_{b,c} \bbk{\epsilon_b\epsilon_c}
  &\approx \f1{4n} \sum_{f,M}
  \sp{\sum_{b_f,c_f} \bbk{\epsilon_{b_f}^f\p{M} \epsilon_{c_f}^f\p{M}}}
  \sp{\sum_{b_{\bar f},c_{\bar f}}
    p_{b_{\bar f}}^{\bar f}\p{M} p_{c_{\bar f}}^{\bar f}\p{M}} \\
  &\approx \f1{4n} \sum_{f,M} \sp{a^f\p{I}
    - \p{1-\delta_{Q^f_\o,0}} a^f\p{M}^2
    - 2\delta_{Q^f_\o,0}} a^{\bar f}\p{M}^2.
\end{align}
Substituting
\begin{align}
  a^1\p{I} = 1,
  &&
  a^2\p{M} = 2\delta_{M,I},
  &&
  \delta_{Q^f_\o,0} = \delta_{f,2},
  &&
  1-\delta_{Q^f_\o,0} = \delta_{f,1},
\end{align}
we thus find that
\begin{align}
  \sum_{b,c} \bbk{\epsilon_b\epsilon_c} = 0.
\end{align}
The infidelity $\bbk{\I}$ is therefore determined entirely by the variances $\bbk{\epsilon_b^2}$.

\subsubsection{Variances}

We now consider the variances
\begin{align}
  \bbk{\epsilon_b^2}
  \approx \f14 \sum_{f,M}
  \bbk{\epsilon^f_{b_f}\p{M}^2} \, p^{\bar f}_{b_{\bar f}}\p{M}^2
\end{align}
where from Eq.~\eqref{eq:cov_frag} we know that
\begin{align}
  n\, \bbk{\epsilon_b^f\p{M}^2}
  = p^f_b\p{I} - p^f_b\p{M}^2\p{1-\delta_{Q^f_\o,0}}
  - \delta_{Q^f_\o,0} \sum_r q^f_b\p{M_r}^2.
  \label{eq:var_frag}
\end{align}
In principle, we have to simplify
\begin{align}
  \bbk{\I} \approx \f14 \sum_b \f{\bbk{\epsilon_b^2}}{p_b},
  &&
  p_b = \frac12 \sum_M p_{b_1}^1\p{M} p_{b_2}^2\p{M},
\end{align}
but this calculation is intractable due to the sum over $M\in\B$ in the denominator.
We therefore settle for trying to find an upper bound on this sum, to which end we observe that
\begin{align}
  n\, \bbk{\epsilon_b^2} \lesssim
  \f14 \sum_{f,M} p^f_{b_f}\p{I} p^{\bar f}_{b_{\bar f}}\p{M}^2
  < \sum_f p^f_{b_f}\p{I} p^{\bar f}_{b_{\bar f}}\p{I}^2.
\end{align}
Rather than the factors $p^f_{b_f}\p{I}$, we can express this bound in terms of the mean probability $\bk{p^f_{b_f}}$ to get bitstring $b_f$ on fragment $f$, averaged over all possible conditions.
This mean probability is $\bk{p^f_{b_f}}=p^f_{b_f}\p{I}/2^{Q^f_\i}$, where $Q^f_\i$ is the number of quantum inputs (initialization conditions) on fragment $f$, so
\begin{align}
  n\, \bbk{\epsilon_b^2}
  \lesssim \sum_f 2^{Q^f_\i} \bk{p^f_{b_f}} \,
  4^{Q^{\bar f}_\i} \bk{p^{\bar f}_{b_{\bar f}}}^2
  = 4 \, \bk{p^1_{b_1}} \bk{p^2_{b_2}}
  \sum_f \f1{2^{Q^f_\i}} \bk{p^{\bar f}_{b_{\bar f}}}.
\end{align}
In turn, we can bound
\begin{align}
  \bbk{\I} \approx \f14 \sum_b \f{\bbk{\epsilon_b^2}}{p_b}
  \lesssim \f1n \sum_b
  \sp{\f{\bk{p^1_{b_1}} \bk{p^2_{b_2}}}{p_b}}
  \sum_f \f1{2^{Q^f_\i}} \bk{p^{\bar f}_{b_{\bar f}}}.
\end{align}
The factor $\bk{p^1_{b_1}}\bk{p^2_{b_2}}/p_b$ is a measure of the quantum correlation between two fragments: it is equal to 1 if there is no correlation, and is smaller (greater) than 1 if quantum correlations cause constructive (destructive) interference for measurement outcome $b$ on the combined circuit.
In principle, this factor can be made arbitrarily large, but that requires fine tuning, and generally speaking we would expect $\bk{p^1_{b_1}}\bk{p^2_{b_2}}/p_b\sim O(1)$ for random circuits.
We therefore expect that
\begin{align}
  \bbk{\I}
  \sim O\p{\f1n \sum_{b,f} \f1{2^{Q^f_\i}} \bk{p^{\bar f}_{b_{\bar f}}}}
  = O\p{\f1n \sum_f 2^{C^f_\o-Q^f_\i}}.
\end{align}

\subsection{The general case}

We now estimate an upper bound on expected infidelity for the general case of $K$ cuts and $F$ fragments.
Defining the projectors
\begin{align}
  \P_\B = \bigcup_{M\in\B}\set{M_r:r\in\lambda\p{M}},
\end{align}
a fragment with $K_f$ incident cuts is nominally characterized by the conditional distributions
\begin{align}
  p^f\p{\v M} = a^f\p{\v M} q^f\p{\v M},
\end{align}
where $\v M\in\P_\B^{K_f}$, $a^f\p{\v M}=\sum_b p_b^f\p{\v M}$ is the normalization of $p^f\p{\v M}$, and $q^f\p{\v M}$ is a normalized probability distribution.
Similarly to before, the estimator $\tilde a^f\p{\v M}\approx a^f\p{\v M}$ is set to the fraction of times that we observe the measurement conditions in $\v M$ when measuring the corresponding qubits in the appropriate bases and preparing the initialization conditions in $\v M$.
Each entry $\tilde q^f_b\p{\v M}$ of the probability distribution $\tilde q^f_b\p{\v M}\approx q^f_b\p{\v M}$ is set to the fraction of times that we observe bitstring $b$ on the classical output of fragment $f$ with conditions $\v M$.
The estimator $\tilde p^f$ and errors $\beta^f,\gamma^f,\epsilon^f$ are defined just as before, and all appropriate objects are are defined for `conditions' $\v M\in\B^{K_f}$ by summing over the spectrum of each condition similarly to Eq.~\eqref{eq:M_cond}.
Altogether, we have the estimator
\begin{align}
  \tilde p = \f1{2^K} \sum_{\v M\in\B^K}
  \bigotimes_f \tilde p^f\p{\v M^f}
  = p + \epsilon,
\end{align}
where $\v M^f\subset\v M$ denotes the conditions in $\v M$ that are incident on fragment $f$, and the distribution error $\epsilon$ has components
\begin{align}
  \epsilon_b \approx \f1{2^K} \sum_{\v M\in\B^K} \sum_f
  \epsilon_{b_f}^f\p{\v M^f} \prod_{g\ne f} p_{b_g}^f\p{\v M^g},
\end{align}
where we neglect terms that are second order or higher in the fragment errors $\epsilon^f$.

\subsubsection{Covariances}

By the same argument as before, the mean $\bbk{\epsilon_b}=0$, so we turn to considering the covariance
\begin{align}
  \bbk{\epsilon_b\epsilon_c}
  \approx \f1{4^K} \sum_{\v M\in\B^K}
  \sum_f \Bbk{\epsilon^f_{b_f}\p{\v M^f}\epsilon^f_{c_f}\p{\v M^f}}
  \prod_{g\ne f} p^g_{b_g}\p{\v M^g} p^g_{c_g}\p{\v M^g},
\end{align}
which leads us to expand
\begin{align}
  \begin{split}
    \Bbk{\epsilon^f_b\p{\v M^f}\epsilon^f_c\p{\v M^f}}
    &\approx \sum_{R,S\in\lambda\p{\v M^f}} \norm{R} \norm{S} \,
    \Bbk{\beta^f\p{\v M^f_R} \beta^f\p{\v M^f_S}}
    q^f_b\p{\v M^f_R} q^f_c\p{\v M^f_S} \\
    &\quad+ \sum_{R,S\in\lambda\p{\v M^f}} \norm{R} \norm{S} \,
    a^f\p{\v M^f_R} a^f\p{\v M^f_S}
    \Bbk{\gamma_b^f\p{\v M^f_R} \gamma_c^f\p{\v M^f_S}},
  \end{split}
\end{align}
where $R=\p{r_1,r_2,\cdots,r_{K_f}}\in\lambda\p{\v M^f}$ is a choice of eigenvalue $r\in\lambda\p{M}$ for each condition $M$ in $\v M^f$, $\v M^f_R$ is the corresponding list of (projectors onto) eigenvectors, and $\norm{R}=\prod_{r\in R} r$.
The covariances between $\beta^f$ and $\gamma^f$ are determined by the multinomial distributions
\begin{align}
  \Bbk{\beta^f\p{\v M^f_R} \beta^f\p{\v M^f_S}}
  &= \f1n \times a^f\p{\v M^f_R}\sp{\delta_{RS} - a^f\p{\v M^f_S}}
  \p{1-\delta_{Q^f_\o,0}}, \\
  \Bbk{\gamma_b^f\p{\v M^f_R} \gamma_c^f\p{\v M^f_S}}
  &= \f1n \times \f{\delta_{RS}}{a^f\p{\v M_R}}
  \times q^f_b\p{\v M^f_R} \sp{\delta_{bc} - q^f_c\p{\v M^f_R}},
\end{align}
which as before implies that
\begin{multline}
  n\, \Bbk{\epsilon^f_b\p{\v M^f}\epsilon^f_c\p{\v M^f}} \\
  \approx \delta_{bc} \, p^f_b\p{\v I^f}
  - \p{1-\delta_{Q^f_\o,0}} p_b^f\p{\v M^f} p_c^f\p{\v M^f}
  - \delta_{Q^f_\o,0} \sum_R p_b^f\p{\v M^f_R} p_c^f\p{\v M^f_R},
\end{multline}
where $\v I^f=\p{I,I,\cdots}$ is a constant list of length $K_f$.
Altogether,
\begin{align}
  \sum_{b,c} \bbk{\epsilon_b\epsilon_b}
  &\approx \f1{4^K n} \sum_{f,\v M} \sum_{b,c}
    \Bbk{\epsilon^f_{b_f}\p{\v M^f} \epsilon^f_{c_f}\p{\v M^f}}
  \prod_{g\ne f} p^g_{b_g}\p{\v M^g} p^g_{c_g}\p{\v M^g} \\
  &\approx \f1{4^K n} \sum_{f,\v M}
  \sp{a^f\p{\v I^f} - \p{1-\delta_{Q^f_\o,0}} a^f\p{\v M^f}^2
  - 2^{K_f} \delta_{Q^f_\o,0}}
  \prod_{g\ne f} a^g\p{\v M^g}^2.
\end{align}
Though unsure how to evaluate this quantity, we can use the fact that $a^f\p{\v M^f}\le a^f\p{\v I^f}^2\le 4^{K_f}$ to bound
\begin{align}
  \f1{4^K} \sum_{f,\v M} a^f\p{\v M^f}^2 \prod_{g\ne f} a^g\p{\v M^g}^2
  < F \prod_g a^g\p{\v I^g}^2
  < 4^K F,
\end{align}
and similarly
\begin{align}
  \f1{4^K} \sum_{f,\v M} 2^{K_f} \prod_{g\ne f} a^g\p{\v M^g}^2
  \le \sum_f 2^{K_f} \prod_{g\ne f} a^g\p{\v I^g}^2
  \le \sum_f 2^{K_f} \prod_{g\ne f} 4^{K_g}
  < 4^K F,
\end{align}
which implies that the contribution to $\bbk{\I}$ from the covariances $\bbk{\epsilon_b\epsilon_b}$ satisfies
\begin{align}
  -\f14\sum_{b,c} \bbk{\epsilon_b\epsilon_b} < \f{4^K F}{4n}.
\end{align}

\subsubsection{Variances}

As in the case of one cut and two fragments, we now bound
\begin{align}
  n \, \bbk{\epsilon_b^2}
  &\lesssim \f1{4^K} \sum_{f,\v M}
  p^f_{b_f}\p{\v I^f} \prod_{g\ne f} p^g_{b_g}\p{\v M^g}^2 \\
  &< \sum_f p^f_{b_f}\p{\v I^f} \prod_{g\ne f} p^g_{b_g}\p{\v I^g}^2
  \label{eq:approx_I} \\
  &= \sp{\prod_h p^h_{b_h}\p{\v I^h}}
  \sum_f \prod_{g\ne f} p^g_{b_g}\p{\v I^g} \\
  &= \sp{\prod_h 2^{Q^h_\i} \bk{p^h_{b_h}}}
  \sum_f \prod_{g\ne f} 2^{Q^g_\i} \bk{p^h_{b_g}} \\
  &= 4^K \sp{\prod_h \bk{p^h_{b_h}}}
  \sum_f 2^{-Q^f_\i} \prod_{g\ne f} \bk{p^h_{b_g}}.
\end{align}
The contribution to infidelity $\bbk{\I}$ from the variances $\bbk{\epsilon_b^2}$ is then
\begin{align}
  \f14 \sum_b \f{\bbk{\epsilon_b^2}}{p_b}
  \lesssim \f{4^K}{4n}
  \sum_b \sp{\f{\prod_h\bk{p^h_{b_h}}}{p_b}}
  \sum_f 2^{-Q^f_\i} \prod_{g\ne f} \bk{p^h_{b_g}}
  \sim O\p{\f{4^K}{n} \sum_f 2^{C^f_\o-Q^f_\i}},
\end{align}
so altogether
\begin{align}
  \bbk{\I} \sim O\p{\f{4^K}{n} \sum_f 2^{C^f_\o-Q^f_\i}}.
  \label{eq:cut_bound}
\end{align}
In practice, we find this asymptotic bound to be overly pessimistic with regards to the scaling with $K$.
There are other ways in which this bound is too {\it optimistic}: by assuming that fragments are weakly correlated and $\frac1{p_b}\prod_h\bk{p^h_{b_h}}\sim O(1)$, this bound does not capture the effect of errors due to noisy virtual teleportation of qubits across cuts between fragments, at their quantum inputs and outputs.
Nonetheless, the bound in Eq.~\eqref{eq:cut_bound} demonstrates $\sim\frac1n\sum_f2^{C^f_\o}$ scaling with shot number $n$ and fragment size $C^f_\o$, which (all else equal) are not affected by these considerations.

\section{Standard deviation of infidelities}
\label{sec:infidelities_std}

\begin{figure}
  \centering
  \includegraphics{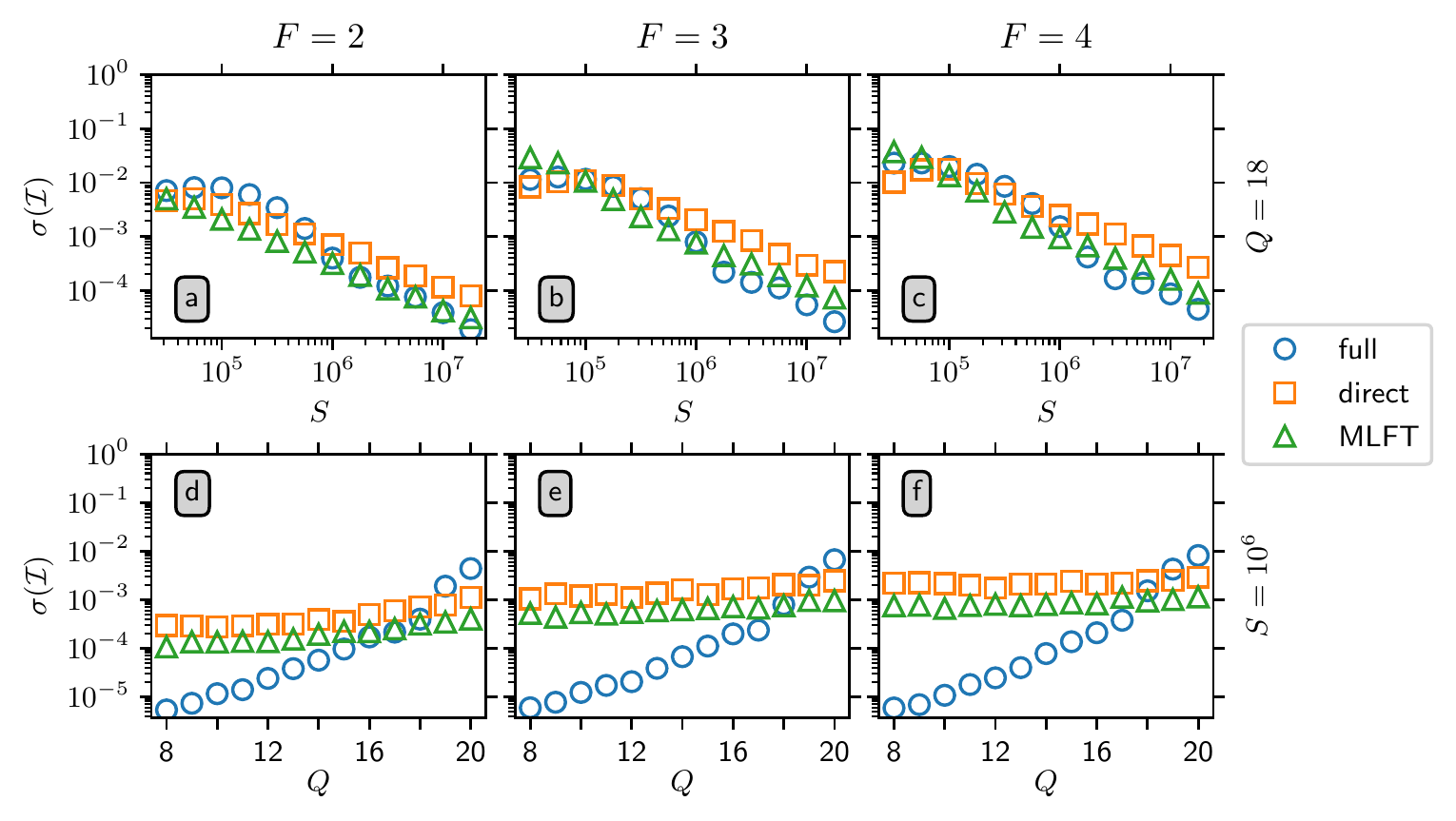}
  \caption{Standard deviation of the infidelities shown in Figure \ref{fig:infidelities}.}
  \label{fig:infidelities_std}
\end{figure}

Figure \ref{fig:infidelities} of the main text shows infidelities $\bbk{\I}$ of reconstructed outputs for clustered random unitary circuits (RUCs).
To provide a sense of the robustness of $\I$ to circuit variations, Figure \ref{fig:infidelities_std} shows the standard deviation $\sigma\p\I=\sqrt{\Bbk{\p{\I-\bbk\I}^2}}$ in the same simulations.
Generally speaking, $\sigma\p\I$ is orders of magnitude smaller than $\bbk{\I}$.

\end{document}